# Analyses of a Datable Solar Eclipse Record in Maya Classic Period Monumental Inscriptions

Hisashi Hayakawa (1-4)*, Misturu Sôma (5), J. Hutch Kinsman (6)

(1) Institute for Space-Earth Environmental Research, Nagoya University, Nagoya, 4648601, Japan

(2) Institute for Advanced Researches, Nagoya University, Nagoya, 4648601, Japan

(3) Science and Technology Facilities Council, RAL Space, Rutherford Appleton Laboratory, Harwell Campus, Didcot, OX11 0QX, UK

(4) Nishina Centre, Riken, Wako, 3510198, Japan

(5) National Astronomical Observatory of Japan, Mitaka, 1818588, Japan

(6) Chesla Dr, Gainesville, Georgia, 30506, USA

* hisashi@nagoya-u.jp

**Abstract**

Historical records of total solar eclipses provide vital information for computing the rotation of the Earth and understanding its long-term variations by providing data from the time before the modern measurements. While eclipses recorded around Eurasia and North Africa for millennia have been subjected to consideration in this context, eclipse records in the American continents have received little attention. In this study, we analysed the solitary observational record for a solar eclipse conducted by the ancient Maya on 16 July 790 in Julian calendar, recorded on the Stela 3 of Santa Elena Poco Uinic (N16°35′, W91°44′). This stela has an eclipse glyph and is associated with a total solar eclipse. Taking the up-to-date Earth rotation ($\Delta T$) rate into account, our calculations locate this site slightly out of the totality path. The visibility of the total solar eclipse from Santa Elena Poco Uinic would require $\Delta T$: 4074 s < $\Delta T$ < 4873 s. In comparison with the contemporary eclipse records, this yields a short-term increase in $\Delta T \geq 800$ s between 761 and 790 and a decrease in $\Delta T \geq 300$ s till to 873. Therefore, the total solar eclipse on 16 July 790 cannot be expected to have been visible from Santa Elena Poco Uinic, unlike what has been previously considered. We conclude that this stela probably records a partial solar eclipse of great magnitude ($\approx 0.946$) visible under favourable meteorological conditions or is based on hearsay from the southern coastal area.

**1. Introduction**

Total solar eclipses are spectacular astronomical events and have been recorded in various civilisations for millennia (*e.g.*, Stephenson, 1997; De Jong and Van Soldt, 1989; Pasachoff and Olson, 2014). These events also provide opportunities observations that can aid scientific endeavours such as spectral analyses, calculations of variable coronal dynamics, and measurements of the variable rotation of the Earth (*e.g.*, Sôma and Tanikawa, 2015; Riley *et al.*, 2015; Stephenson *et al.*, 2016; Pasachoff, 2017; Morrison *et al.*, 2019; Hayakawa *et al.*, 2020, 2021). Visibility of total solar eclipses requires an exact apparent coincidence of the solar disk and the lunar disk. This condition forms a narrow path of totality within a wider zone where a partial eclipse is visible.

Thus, total solar eclipses provide vital information for the parameter $\Delta T$ related to the long-term variability of the Earth's rotation rate; $\Delta T$ is defined as the difference between a uniform timescale (terrestrial time = TT) and a timescale affected by the Earth's rotation (universal time = UT). Eclipse records have been surveyed on a millennial timescale (Stephenson, 1997; Tanikawa and Sôma, 2004; Sôma and Tanikawa, 2015; Stephenson *et al.*, 2016; Morrison *et al.*, 2019) in combination with historical occultation records (Sôma and Tanikawa, 2016; Gonzalez, 2017, 2019) to reconstruct the long-term $\Delta T$ variability, especially for the period before the systematic measurements with atomic time scales since the last century and observations of timed occultations since the 17th century (*e.g.*, Calame and Mulholland, 1978; Morrison *et al.*, 1981; Herald and Gault, 2012).

While records from various regions in Eurasia have been studied (*e.g.*, Stephenson, 1997; Stephenson *et al.*, 2016), no reports from the American continents have been exploited yet in this regard, despite the significant astronomical achievements of the ancient American civilizations. The Maya, with their highly developed calendar system (*e.g.*, Teeple, 1931; Thompson, 1950; Kelley, 1976; Bricker and Bricker, 2011; Aldana y Villalobos and Barnhart, 2014) and extensive compilations of canonical eclipses in tables found in the Dresden Codex (Thompson, 1972; Bricker and Bricker, 2011; Aldana y Villalobos and Barnhart, 2014; *c.f.*, Love, 2017), are particularly notable in this regard. In this context, we analyse a Maya eclipse stela that has been associated with a total solar eclipse in 790 (*e.g.*, Palacios, 1928; Teeple, 1931; Bricker and Bricker, 2011; Love, 2017), in comparison with the long-term $\Delta T$ variability (*e.g.*, Stephenson *et al.*, 2016).

**2. Stela 3 of Santa Elena Poco Uinic**

Among the plethora of Maya stone monuments in the Classic Maya in the Maya Classic Period (AD 250~900), a unique record of a solar eclipse was found at the site of Santa Elena Poco Uinic in Chiapas, Mexico, while some more inscriptions have been tentatively associated with solar eclipses (*e.g.*, Bricker and Bricker, 2011, p. 364). Santa Elena Poco Uinic is situated in the canyon region of Chiapas described as "spectacular" by Mathews[1], located in eastern Chiapas (N16°35′, W 91°44′) southward from Palenque (Prager *et al*., 2014), and was possibly the capital of a small kingdom. Carved into the stone monolith known as Stela 3 (Figure 1), the aforementioned inscription records observations on a solitary solar eclipse in the Maya civilisation with a glyph that is unique to solar eclipses (Teeple, 1931, p. 115; Thompson, 1935, p. 74; Martin and Skidmore, 2012, p. 6; Love, 2017; Kinsman and Asher, 2017). It is dated "5 Kib 14 Ch'en" and long count "9.17.19.13.16" in the Maya calendar (Mathews, 2006), which converts to 16 July AD 790 in the Julian calendar under the Martin–Skidmore correlation (Martin and Skidmore, 2012; Kennett *et al*., 2013; Iwaniszewski, 2021). Studies have used the direct association of the Maya eclipse date with the actual solar eclipse on this date to derive the latest correlation between the Maya and Christian calendars, one of popular conversions within the Maya studies (Martin and Skidmore, 2012). Stela 3 was likely sculpted and commemorated on 8 October 790.

---

[1] http://research.famsi.org/whos_who/sites.php?sitename=Santa%20Elena%20Poco%20Uinic

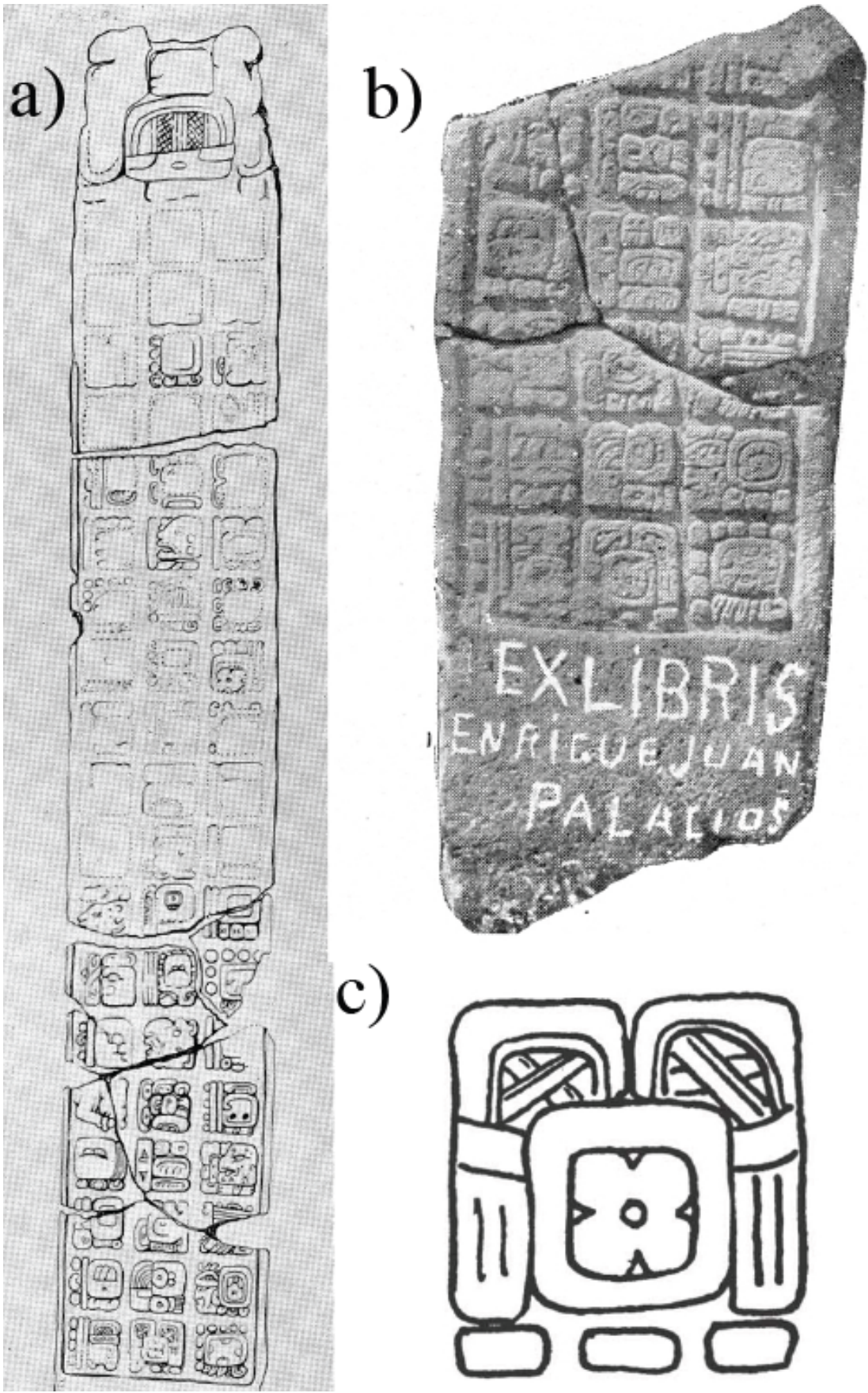

Figure 1: Eclipse glyph on Stela 3 of Santa Elena Poco Uinic. Panel a) shows L. Orellana's overall trace copy for the Stela 3 in Palacios (1928, p. 124). Panel b) shows M. O. de Mendizabal and F. Tannenbaum's photograph for its bottom part in Palacios (1928, p. 211). Panel c) shows J. E. Teeple's close-up trace for the eclipse glyph in Teeple (1931, p. 115).

The eclipse glyph appears as the central hieroglyph in the bottom-most row of three hieroglyphs on Stela 3 and represents the event on the date immediately preceding this glyph. This hieroglyph has been iconographically interpreted as denoting the "disappearance of the sun" and associated with the total solar eclipse on this date (Teeple, 1931, p. 115; see also Love, 2017, p. 3). For this eclipse, Bricker and Bricker (2011, p. 364) stated: "A total solar eclipse affected the entire Maya area three days later, on 20 July AD 790 [NB in Gregorian calendar] at about 1:00 p.m., with the path of centrality passing directly over Chiapas", using another calendar conversion. While this interpretation is widely accepted by Maya studies scholars, its phonetic reading in the parlance of hieroglyphic writing is controversial. The *k'in*, or Sun sign, similar to a four-leaf clover, is in the centre of the hieroglyph flanked by two cross-band signs, which are possibly sky signs (Martin and Skidmore, 2012, p. 6). These twin wing-like signs are similar to the flanking motifs found in the Dresden Codex in what some scholars call "eclipse" glyphs (*e.g.*, Bricker and Bricker, 2011; Prager, 2006). However, most recently, Love (2017) argued against the connection of these eclipse "look-a-likes"' to the hieroglyph found at Poco Uinic. Further, since the eclipse glyph on Stela 3 is unique in the corpus of Maya hieroglyphic inscriptions, despite many similar glyphs purported to refer to eclipses (Martin and Skidmore, 2012, p. 6, Figures 3 and 4), a viable phonetic reading for the glyph is challenging.

### 3. Eclipse Visibility and Earth's Rotation Rate

In contrast, our calculations with the up-to-date variable Earth rotation locates Poco Uinic slightly out of the totality path. To compute the totality path, we calculated the coordinates of the Sun and Moon with JPL ephemerides DE431 (Folkner *et al*., 2014) and applied the existing $\Delta T \approx 2900$ s in 790 based on Stephenson *et al*. (2016). The results of our calculations are demonstrated in Figure 2, which shows the eclipse's totality path passing off Santa Elena Poco Uinic. From this location, a partial eclipse was visible during 11:47–14:59 LT (apparent solar time), and the eclipse magnitude reached only up to 0.946 at this location, even at its maximum at 13:29 LT (Figure 3). This contradicts slightly the notion that this stela records the visible total eclipse on 16 July 790 (*e.g.*, Teeple, 1931; Martin and Skidmore, 2012; Love, 2017). In this case, it was probably challenging to see the solar disk during partial eclipse without appropriate filters (*e.g.*, Chou, 2016).

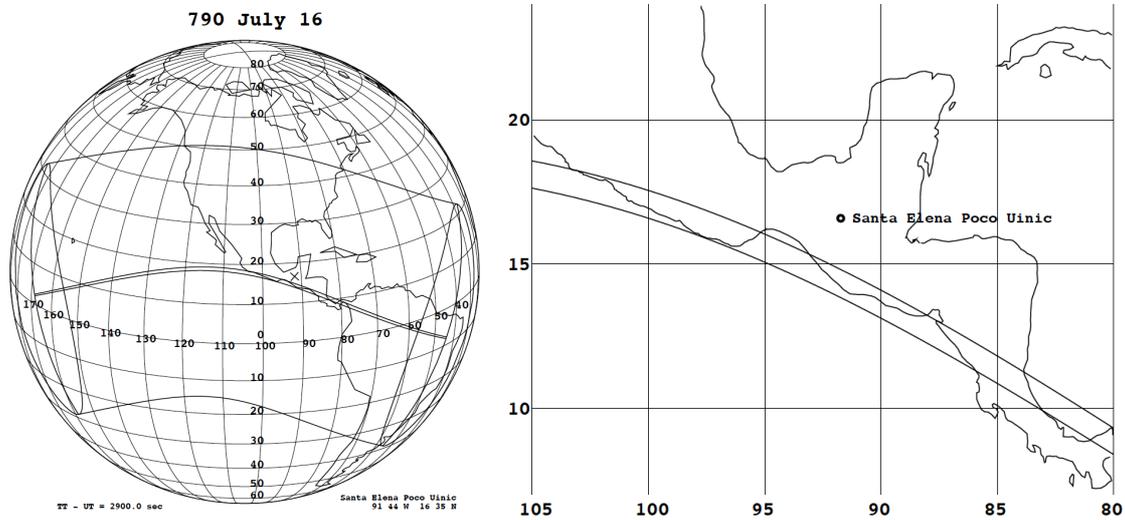

Figure 2: Totality path of solar eclipse on 16 July 790 (a) and its close-up around Santa Elena Poco Uinic (b), calculated assuming ΔT ≈ 2900 s in 790 based on Stephenson *et al*. (2016). These figures show that the total eclipse was seen only along the Pacific coast of Mesoamerica, whereas partial eclipse was seen widely in Mesoamerica.

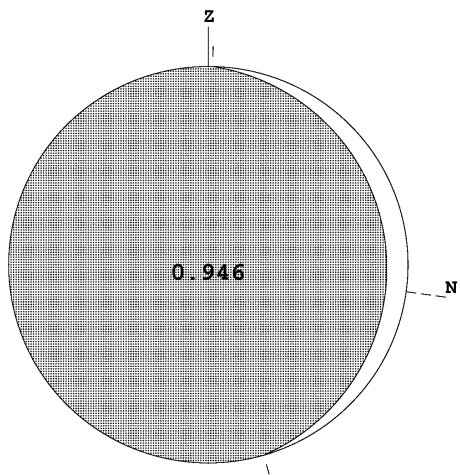

Figure 3: Eclipse seen at Santa Elena Poco Uinic at highest magnitude (0.946) at 13:29 LT, calculated with the ΔT ≈ 2900 s in 790 based on Stephenson *et al*. (2016).

In contrast, following the existing total-eclipse interpretations (Teeple, 1931, p. 115; Love, 2017, p. 3), we need to locate Santa Elena Poco Uinic within the totality path. In order to calculate the location of the total or annular eclipse path, one needs the value of ΔT. First, we assume that ΔT=0s and calculate the location of the northern and southern limits of the eclipse band following the procedures given in Explanatory Supplement (H. M. Nautical Almanac Office, 1961, pp. 224-227). Let the geodetic longitude and latitude of the position where the total or annular eclipse was observed be λ and φ, and calculate the range of longitude $λ_1$ to $λ_2$ where the total or annular eclipse was seen for the latitude φ. Then, the range of ΔT can be obtained from the following formula, where λ, $λ_1$, and $λ_2$ are expressed in degrees:

$$1.0027379 \times 240s\ (λ － λ_2) < ΔT < 1.0027379 \times 240s\ (λ － λ_1),$$

Accordingly, its visibility at Santa Elena Poco Uinic in 790 requires the ΔT as 4074 s < ΔT < 4873 s. This is significantly larger than that in Stephenson *et al.*'s estimate (ΔT ≈ 2900 s). Simultaneously, we note that records from the period around this eclipse are scarce. As shown in Figures 12 and 15 in Stephenson *et al.* (2016), three eclipse records have been used to constrain the ΔT variation between 700 and 900: those on 5 August 761 at Cháng'ān (1720 s < ΔT < 3290 s), 5 May 840 at Bergamo (1610 s < ΔT < 6800 s), and 28 July 873 at Nīšāpūr (1820 s < ΔT < 3750 s) (Stephenson, 1997; Stephenson *et al.*, 2016). Among these, the 873 eclipse has been interpreted to imply ΔT ≈ 2300 s in Stephenson *et al.* (2016), which has been accompanied by an annular eclipse record at Kyoto (N 35°01′, E 135°45′) whose ΔT range is computed to be constrained to 3327 s < ΔT < 3498 s to allow simultaneous observations at Nīšāpūr and Kyoto and other contemporary observations (Tanikawa and Sôma, 2004). Accordingly, the visibility of the total eclipse at Santa Elena Poco Uinic would require a short-term increase in ΔT ≥ 800 s between 761 and 790 and a decrease in ΔT ≥ 300 s to 873 (see Figure 4). The former is unrealistic, unless otherwise supported by robust reports of contemporary total solar eclipses.

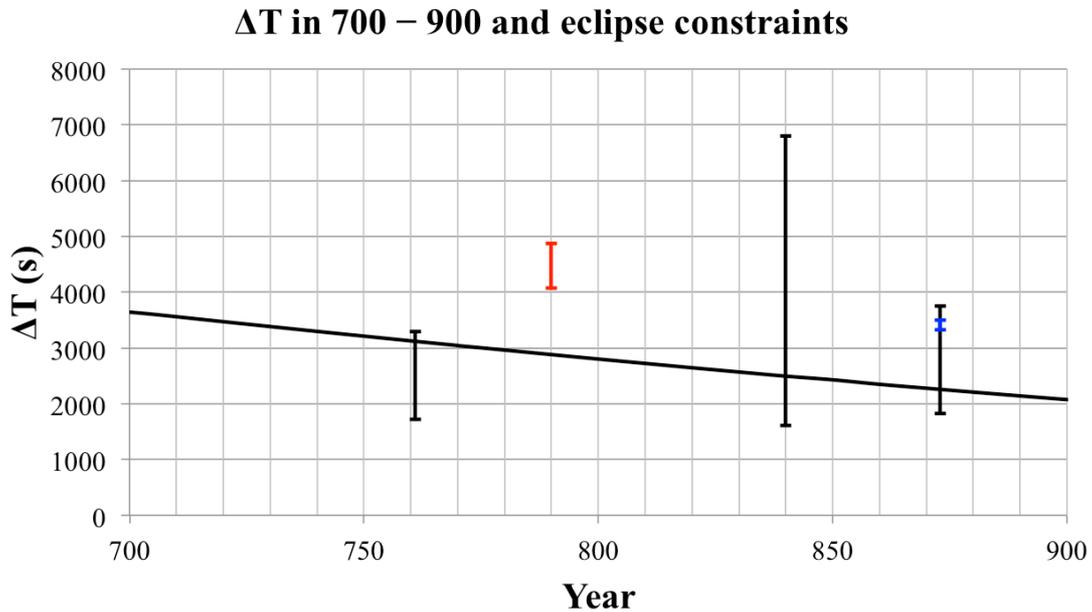

Figure 4: Temporal variation of ΔT and constraints with eclipse records from Stephenson *et al.* (2016) in black bars and from Tanikawa and Sôma (2004) in blue bar, as well as the ΔT range required to realise total eclipse at Santa Elena Poco Uinic on 16 July 790 in red bar.

**4. Summary and Discussions**

In this study, we analysed the solitary Maya observation of a solar eclipse on Stela 3 of Santa Elena Poco Uinic, which has been interpreted to mean the "disappearance of the sun" on 16 July 790, indicating the observation the total eclipse from its site (N 16°35′, W 91°44′). While its iconographic interpretation of a solar eclipse seems robust, the phonetic interpretation is still controversial.

In contrast, our calculation with the latest ΔT (≈ 2900 s in 790; Stephenson *et al.*, 2016) concluded that this site was off the totality path of this eclipse and that only a partial eclipse was visible from there. The maximum magnitude of the partial eclipse is calculated as 0.946 at 13:29 LT during its period of visibility: 11:47–14:59 LT on 16 July 760. For the total solar eclipse to have been visible from this site, ΔT in that year should have to be significantly larger (4074 s < ΔT < 4873 s) than the current estimate of ΔT ≈ 2900 s (Stephenson *et al.*, 2016). In comparison with the existing constraints in Stephenson *et al.* (2016), this would require a short-term increase in ΔT ≥ 800 s between 761 and 790, making the visibility of the total eclipse from this site (Teeple, 1931; Love, 2017) unlikely.

While naked-eye observations of partial eclipses are challenging even at high magnitudes (*e.g.*,

Chou, 2016), several possibilities have been suggested. First, the Maya could have seen this deep partial eclipse under favourable meteorological conditions, such as a thin cloud cover as a natural filter. Actual cases of naked-eye sunspot observations through cloud cover (*e.g.*, Willis *et al.*, 2018; Hayakawa *et al.*, 2019) have demonstrated that the solar disk can be visible under favourable meteorological conditions. In this case, given the great eclipse magnitude (0.946 at its maximum), it would most likely have been noticeable from Santa Elena Poco Uinic through suitable natural filters. As the Chiapas region in July would be in the middle of a rainy season, such a thin cloud cover is something that could be easily expected, even in the present times. The forest canopy was obviously much denser in the Classic period. Alternatively, residents of Santa Elena Poco Uinic may have received reports of the total eclipse from their neighbours in the southern territory, as extensions of the Maya settlements in the Classic Period are confirmed to have extended southward (*e.g.*, Figure 1 of Taladoire, 2015). In this case, these reports would have been brought from the southern territory of the current Guatemala.

Therefore, our analyses confirm that the solar eclipse of Santa Elena Poco Uinic Stela 3 was indeed astronomically observable from the said site on 16 July 790, not as a total eclipse but as a deep partial eclipse, in contrast with what has been previously considered (*e.g.*, Bricker and Bricker, 2011, p. 364). Given current known evidence, visibility of the total eclipse from this site would necessitate a short-term increase in $\Delta T \geq 800$ s between 761 and 790, which is slightly unrealistic. This makes observations of great partial eclipse under favourable meteorological conditions or hearsays of the totality from the southern sites more plausible compared to a total eclipse observation.

**Acknowledgments**

We thank Barbara MacLeod and Carlos Pallan for their helpful advice and discussions on the philological interpretations. HH thanks Masaki Kuwada for her hospitality during his fieldwork and Yuri I. Fujii for her help on accessing Palacios (1928). HH also thanks Gioacchino Falsone, Marshall Becker, Anita Fahringer, So Miyagawa, and Yukiko Kawamoto for their advice. This work has been supported with JSPS Grant-in-Aids JP20K22367 and 20H05643, JSPS Overseas Challenge Program for Young Researchers, Young Leader Cultivation (YLC) program of Nagoya University, the YLC collaborating research funds in 2020 – 2021, and the research grants for Mission Research on Sustainable Humanosphere from Research Institute for Sustainable Humanosphere (RISH) of Kyoto University.

Hayakawa et al. (2021) Analyses of the Maya Eclipse Record, *Publications of the Astronomical Society of Japan*, DOI: 10.1093/pasj/psab088

**Data Availability**

The original Maya stela is situated at Santa Elena Poco Uinic. We have consulted its photograph and copies in Palacios (1926). We have computed the coordinates of the Sun and Moon with NASA JPL ephemerides DE431 (Folkner *et al.*, 2014) and derived the ΔT variability from Stephenson *et al.* (2016).

**References**

Aldana y Villalobos, G., Barnhart, E. L.: 2014, *Archaeoastronomy and the Maya*, Oxbowbooks, Oxford.

Bricker, H. M., Bricker, V. R.: 2011, *Astronomy in the Maya Codices*, Philadelphia: American Philosophical Society.

Calame, O., Mulholland, J. D.: 1978, *Science*, **199**, 977-978. DOI: 10.1126/science.199.4332.977

Chou, R.: 2016, *American Astronomical Society*, **2016**, 1-9.

De Jong, T., Van Soldt, W. H.: 1989, *Nature*, **338**, 238–240. DOI: 10.1038/338238a0

Folkner, W. M., Williams, J. G., Boggs, D. H., Park, R. S., Kuchynka, P.: 2014, *IPN Progress Report*, **2014-02-15**, 42-196.

Gonzalez, G.: 2017, *Monthly Notices of the Royal Astronomical Society*, **470**, 1436-1441. DOI: 10.1093/mnras/stx1316

Gonzalez, G.: 2019, *Monthly Notices of the Royal Astronomical Society*, **482**, 1452-1455. DOI: 10.1093/mnras/sty2820

Hayakawa, H., Willis, D. M., Hattori, K., Notsu, Y., Wild, M. N., Karoff, C.: 2019, *Solar Physics*, **294**, 95. DOI: 10.1007/s11207-019-1488-5

Hayakawa, H., Owens, M. J., Lockwood, M., Sôma, M.: 2020, *The Astrophysical Journal*, **900**, 114. DOI: 10.3847/1538-4357/ab9807

Hayakawa, H., Lockwood, M., Owens, M. J., Sôma, M., Besser, B. P., Van Driel, L.: 2021, *Journal of Space Weather and Space Climate*, **11**, 1. DOI: 10.1051/swsc/2020035

Herald, D., Gault, D.: 2012, *International Occultation Timing Association (IOTA)*, https://ui.adsabs.harvard.edu/abs/2012yCat.6132....0H/abstract

H.M. Nautical Almanac Office, 1961, *Explanatory Supplement to the Astronomical Ephemeris and the American Ephemeris and Nautical Almanac*, London, Her Majesty's Stationery Office.

Iwaniszewski, S.: 2021, Remarks on the Lunar Series and Eclipse Cycles in Late Classic Maya Records. In: Boutsikas, E., McCluskey, S. C., Steele, J. (eds.) *Advancing Cultural Astronomy*, Springer, Cham, pp. 237-249. DOI: 10.1007/978-3-030-64606-6_12

Kennett, D. J., Hajdas, I., Culleton, B. J., *et al*.: 2013, *Scientific Reports*, **3**, 1597. DOI: 10.1038/srep01597

Kinsman, J. H., Asher, D. J.: 2017, *Planetary and Space Science*, **144**, 112-125. DOI: 10.1016/j.pss.2017.04.024

Launay, F.: 2012, *The Astronomer Jules Janssen: A Globetrotter of Celestial Physics* (S. Dunlop, tr.), Springer, New York.

Love, B.: 2017, *Ancient Mesoamerica*, **29**, 219-244. DOI: 10.1017/S0956536116000444

Martin, S., Skidmore, J.: 2012, *The PARI Journal*, **13**, 3-16

Morrison, L. V., Lukac, M. R., Stephenson, F. R.: 1981, *Royal Greenwich Observatory Bulletins*, 186.

Morrison, L. V., Stephenson, F. R., Hohenkerk, C. Y.: 2019, *Journal for the History of Astronomy*, **50**, 366-372. DOI: 10.1177/0021828619863243

Oppolzer, T.: 1887, *Canon der Finsternisse*, Wien: Kaiserlich-Königlichen Hof-und Staatsdruckerei

Palacios, E. J.: 1928, *En los confines de la selva lacandona. Exploraciones en el estado de Chiapas. Mayo-Agosto 1926*, México, Talleres Gráficos de la Nación.

Pasachoff, J. M.: 2017, *Nature Astronomy*, **1**, 0190. DOI: 10.1038/s41550-017-0190

Pasachoff, J. M., Olson, R. J. M.: 2014, *Nature*, **508**, 314-315. DOI: 10.1038/508314a

Prager, C., Gronemeyer, S., Wagner, E., Matsumoto, M., Kiel, N.: 2014, *A Checklist of Archaeological Sites with Hieroglyphic Inscriptions*. Published on: https://mayawoerterbuch.de

Prager, C.: 2006, IS T326 *Notes on Ancient Maya Writing*, **2006-12-30**, 1-9.

Riley, P., Lionello, R., Linker, J. A., *et al*.: 2015, *The Astrophysical Journal*, **802**, 105. DOI: 10.1088/0004-637X/802/2/105

Stephenson, F. R., Morrison, L. V., Hohenkerk, C. Y.: 2016, *Proceedings of the Royal Society A*, **472**, 20160404. DOI: 10.1098/rspa.2016.0404

Stephenson, F. R.: 1997, *Historical Eclipses and Earth's Rotation*, Cambridge, Cambridge University Press.

Sôma, M., Tanikawa, K.: 2015, Determination of ΔT and Lunar Tidal Acceleration from Ancient Eclipses and Occultations, in: W. Orchiston et al. (eds.), *New Insights From Recent Studies in Historical Astronomy: Following in the Footsteps of F. Richard Stephenson*, Springer, Cham, pp. 11-23. DOI: 10.1007/978-3-319-07614-0_2

Sôma, M., Tanikawa, K.: 2016, *Publications of the Astronomical Society of Japan*, **68**, 29. DOI: 10.1093/pasj/psw020

Taladoire, E.: 2015, *Estudios de Cultura Maya*, **46**, 45-70.

Tanikawa, K., Sôma, M.: 2004, *Publications of the Astronomical Society of Japan*, **56**, 879-885. DOI: 10.1093/pasj/56.5.879

Teeple, J. E.: 1931, *Contributions to American Archaeology*, **1**, 29-116.

Thompson, J. E. S.: 1935, Contributions to American Archaeology, 3(14):51-104. Publication 456. Carnegie Institution of Washington, Washington, D.C.

Thompson, J. E. S.: 1950, *Maya Hieroglyphic Writing: Introduction.* Publication 589. Carnegie Institution of Washington, Washington, D.C.

Thompson, J. E. S.: 1972, *A Commentary on the Dresden Codex: A Maya Hieroglyphic Book,* American Philosophical Society, Philadelphia.

Willis, D. M., Wilkinson, J., Scott, C. J., *et al*.: 2018, *Space Weather*, **16**, 1740-1752. DOI: 10.1029/2018SW002012